\documentclass[showpacs,twocolumn,prl]{revtex4}
\usepackage{amsmath,graphicx}

\begin{document}

\title{Phase dependent Andreev spectrum in a difusive SNS junction. Static and dynamic current response.}
\author{ M. Ferrier$^{1}$, B. Dassonneville$^{1}$ ,  S. Gu\'eron$^{1}$ and H.Bouchiat$^{1}$ }
\affiliation{$^{1}$ LPS, Univ. Paris-Sud, CNRS, UMR 8502, F-91405 Orsay Cedex, France}%,$^2$  IEF, Univ. Paris-Sud, CNRS, UMR 8502, F-91405 Orsay Cedex}
\begin{abstract}
 A long phase coherent normal (N) wire between superconductors (S) is characterized by a dense phase dependent Andreev spectrum. We investigate the  current response of  Andreev states  of an NS ring to a time dependent Aharonov Bohm flux superimposed to a dc one. The ring is modeled with a tight binding Hamiltonian including a  superconducting region with a  BCS coupling between electron and hole states, in contact with a  normal region with on site disorder. Both dc and ac currents are determined from the computed eigenstates and energies using a Kubo formula approach.  Beside the well known Josephson current we identify different contributions to the ac response. A low frequency one related to the dynamics of the thermal occupations of the Andreev states and a higher frequency one related to  microwave induced transitions between levels. Both are characterized by   phase dependencies, with a high harmonics content,  opposite to one another. Our findings are successfully  compared to the results of recent experiments. %In particular we identify a range of parameters for which the flux dependent dissipative response is simply proportional to the minigap.
\end{abstract}
\maketitle
\section{Introduction}
Most properties of a non superconducting N metal connected to two superconductors (an SNS junction) can be seen as resulting from the  phase dependent Andreev states (AS) in the N metal. These eigenstates   are described by coherent combinations of electron and hole wave functions, determined by boundary conditions imposed by the superconducting contacts \cite{kulik}. Whereas most equilibrium properties of  SNS junctions are well understood theoretically and experimentally \cite{heikkila, illichev, lesueur, strunk}, their  high frequency dynamics is a  more complex issue which has only been addressed very recently via the investigation \cite{chiodi2011,dassonneville} of NS rings submitted to a dc Aharonov Bohm  flux $\Phi_{dc}$ with a small ac modulation $\delta \Phi_{\omega}\exp(-i\omega t)$. The quantity measured is the ac current response $\delta I\omega$ superimposed to the dc Josephson current. Within linear response,  $\delta I_\omega$ is related to $\delta \Phi_{\omega}$ by the complex susceptibility $\chi(\omega)= \delta I\omega/\delta \Phi_{\omega}= i\omega Y$ where $Y$ is the impedance of the NS ring. Our work is motivated by these recent experiments \cite{chiodi2011,dassonneville}  which revealed   the dc flux, frequency and temperature dependences of the   response  function $\chi(\omega)$ and related them to the various relevant energy scales:  the Thouless energy $E_{Th}$(inverse diffusion time through the N wire) and the relaxation rate of the population of the Andreev levels.  On the theoretical side, the linear response of SNS junctions has been investigated using time dependent  Keldysh-Usadel  equations \cite {virtanen}. Whereas a good agreement is found with experimental results in the frequency range dominated by relaxation processes of the population of the Andreev levels, the theoretical results obtained  at higher frequency, i.e. in the regime 
where  the dynamics  is dominated by quasi resonant absorption  of photons do not agree with experimental findings.  In order to elucidate this disagreement  we have performed a Kubo analysis of the linear  current response of an NS ring to an ac flux, calculated from  the Andreev eigenstates and energies.   The diffusive NS ring  is described with a tight binding  Bogoliubov-de Gennes Hamiltonian. As detailed in section II, this Hamiltonian describes a ring containing a superconducting region with a  BCS coupling between electron and hole states, in contact with a  normal region with on site (Anderson type)disorder and a vector potential imposing the phase($\varphi$) dependent boundary condition. The eigenstate spectrum is obtained by numerical diagonalisation.  For a long diffusive N metallic wire  (of length $L$ greater than the superconducting coherence length $\xi_s$),  we find that as expected the spectrum  exhibits a phase dependent gap $2E_g (\varphi)$ \cite{heikkila,spivak}.% proportional to the Thouless energy $E_{Th}=\hbar/\tau_D $ where $\tau_D = L^2/D$ is the diffusion time through the N wire. 
 This so-called minigap, much smaller than the superconducting gap $\Delta$, is fully modulated by the phase difference of the superconducting  order parameter $\varphi$ across the N region. $E_g (\varphi)$ is maximal at $\varphi=0$ with 
 $E_g (0)\simeq 3.1 E_{Th}$ and goes linearly to zero at $\varphi= \pi$, approximatively like $E_g(\varphi) \simeq  E_g (0) |\cos(\varphi/2)|$ \cite{spivak,blatter}.  The phase dependent Josephson current $ I_J(\varphi)$ at equilibrium  is calculated by summing  the contributions  of each AS of energy $\epsilon_n$, via $i_n=-\frac{2e}{\hbar}\frac{ \partial\epsilon_n}{\partial \varphi}$, the current carried by level $n$  of   thermal occupation factor $f_n (\varphi)=f(\epsilon_n(\varphi))$ where $f(\epsilon)$ is the Fermi Dirac distribution function.
 \begin{equation}
   I_J(\varphi) =\sum_n f_n(\varphi) i_n (\varphi)
   \label{equIJ}
 \end{equation}
 In section III we  show how to compute  from   the Andreev levels and eigenstates, the ac linear response of the NS ring to an ac flux,  using a Kubo formula similarly to what was previously done in normal Aharonov Bohm rings \cite{buttiker,trivedi, reulet}. One can identify two main mechanisms responsible for the frequency dependence of the in phase susceptibility and correlatively the existence of an out-of-phase dissipative response.  

The first mechanism, discussed in section IV, is  the relaxation of the thermal populations of the Andreev levels with a time scale $\tau_{in}$,  the inelastic  scattering time. It leads to a response $\chi_D$ that  can be expressed with the diagonal  matrix elements of the current operator.  This mechanism is at the origin of a drastic increase of the harmonics content of the non dissipative response, in contrast with the   zero frequency susceptibility  $\chi(0)=\chi_J=\partial I_J / \partial \Phi$  which is a pure cosine  in the same regime of temperature. The dissipative response $\chi'' _D$ is nearly $\pi$ periodic with extra cusps at  $\pi$ that 
reflect the closing of the minigap. 

The second mechanism, discussed in section V,  dominates at frequencies $\omega \tau_{in}\gg 1$. It corresponds  to  quasi resonant transitions above the minigap within frequency scales of the order of $E_g(\varphi)/\hbar$.  In the limit where $\hbar\omega \gg E_g \gg k_BT$ this  phase  dependent dissipative response is simply proportional to the  opposite of the minigap. In the other limit $ k_BT \gg E_g \gg \hbar\omega$, this dissipative response is  mainly determined by the flux  dependence of the non-diagonal  matrix elements of the current operator and is  reversed in sign compared to the  diagonal ones determining the low frequency phase dependent dissipation $\chi'' _D$ .
 In the conclusion we compare our results to recent experiments \cite{chiodi2011, dassonneville} and  theoretical results based on frequency dependent Usadel equations \cite{virtanen}.

 \section{Tight binding Hamiltonian for a diffusive SNS ring}
  We implement the Bogoliubov-de Gennes Hamiltonian described by the 4 blocks matrix, 
  
  \begin{equation}
  \mathcal{H} = \left( \begin{array}{cc}
H-E_F & {\bf \Delta}\\
{\bf \Delta}& E_F-H^*
\end{array} \right)
\label{bogo}
\end{equation}
  
  where  $H$ and $-H^*$  are $N\times N$ matrices that describe respectively the electron and hole like wave function components of a hybrid  NS ring within a tight binding 2D Anderson model. 
  \begin{equation}
  H= \sum_{i=1}^N \epsilon_i |i ><i|  + \sum _{i\neq j}  t_{ij} |i ><j|  
  \end{equation}
The ring has  N = $ N^N +N^S = N_x\times N_y$  sites on a square lattice of period $a$, with a normal portion of $N^N=N_x^N \times N_y$ sites  in contact with a superconducting one   ($N^S =N_x^S \times N_y $ sites). The on-site  random energies  $\epsilon_i$ of zero average and  variance $W^2$ describe the disorder in the ring. The hopping matrix element between nearest neighbours  reads $t_{ij}=t \exp i\varphi_{ij}$ where the phase factor is related to the superconducting phase difference through the normal junction via:   $\varphi_{ij} = (\pi /2\Phi_0)\int \vec A \vec dl = \varphi (x_i-x_j) /N_x^N $, describes the effect of an Aharonov-Bohm flux  $\Phi = A N_x^N a = \Phi_0\varphi/  2\pi$  and $\Phi_0 =h/2e$ is the superconducting flux quantum. For sites in the S part  $\varphi_{ij}=0$.  The BCS diagonal matrix ${\bf\Delta}$ couples electron and hole states exclusively  in the S part $\Delta_{i,i} = \Delta$ for $N_N +1\leq i \leq N$ and is zero otherwise.  We have chosen the amplitude of the superconducting gap $\Delta =t/4$ such that the S coherence length $\xi_s= a t/\Delta \ll N_x^S$ in order to avoid any reduction of the superconducting correlations in the S region (inverse proximity effect). The number of  transverse channels  and the amplitude of the disorder correspond to  the diffusive regime where the length $N_x a$ of the normal region  is longer than the elastic mean free path $l_e$ and shorter than the localization length $N_y l_e$. The length $l_e$ is related to the amplitude of disorder by $l_e \simeq  a 15 (t/W)^2$ at 2D  \cite{sigetti}.  We checked that the results do not depend on the position of the Fermi energy, typically chosen at filling 1/4.  Hereafter, all energies are taken relatively to $E_F$.
\begin{figure}
\center
    \includegraphics[clip=true,width=9cm]{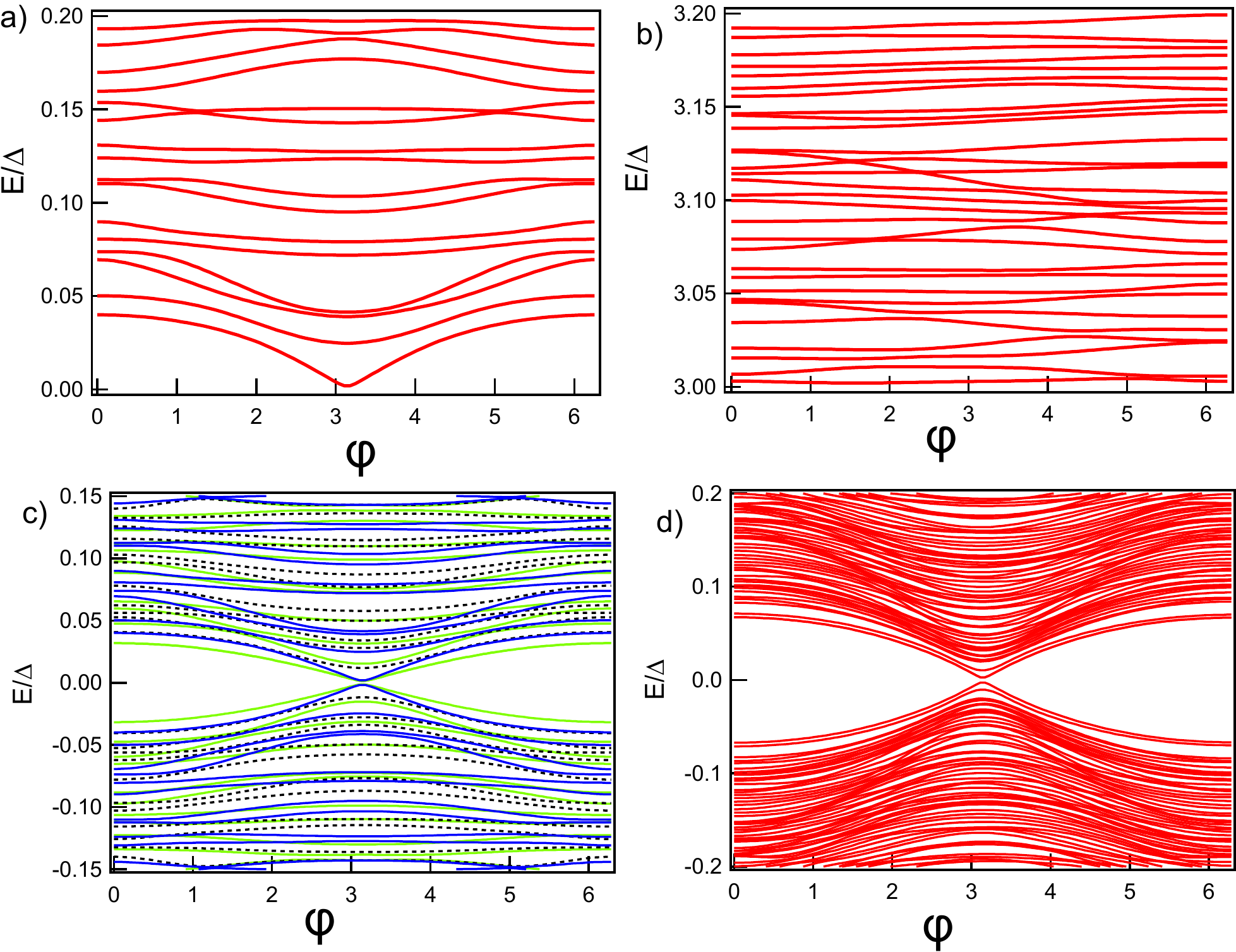}
\caption{Phase dependent spectrum of Andreev levels,  (a): close to the Fermi energy  and (b): above the superconducting gap   for a diffusive ring with  $ N_x^N=  60 \times 24 $ normal sites and on site disorder of amplitude $W/t=1.5$. (The number of S sites with  $\Delta = t/4$ is  
$N^S=50  \times 24$ ), note the denser spectrum above the minigap and the different periodicity.  (c): same as (a), for 3 different disorder configurations (different colors). Note the  symmetry of the spectrum  with respect to the Fermi level at zero energy as well as the opening of the phase dependent minigap which amplitude scales with the Thouless energy, $E_{Th} = \delta_N N_y l_e /N^N a $ where $\delta_N$ is the energy level spacing in the normal region. (d): diffusive ring with  a larger number of transverse channels $ N^N= N_x ^N \times N_y= 20  \times 100 $ normal sites  for a single  disorder configuration of amplitude $W/t=1.5$.(The number of S sites is $N^S=20 \times 100$).}
       \label{spectre}
\end{figure}

\subsection{Minigap and dc Josephson current}
\begin{figure}
    \includegraphics[clip=true,width=9cm]{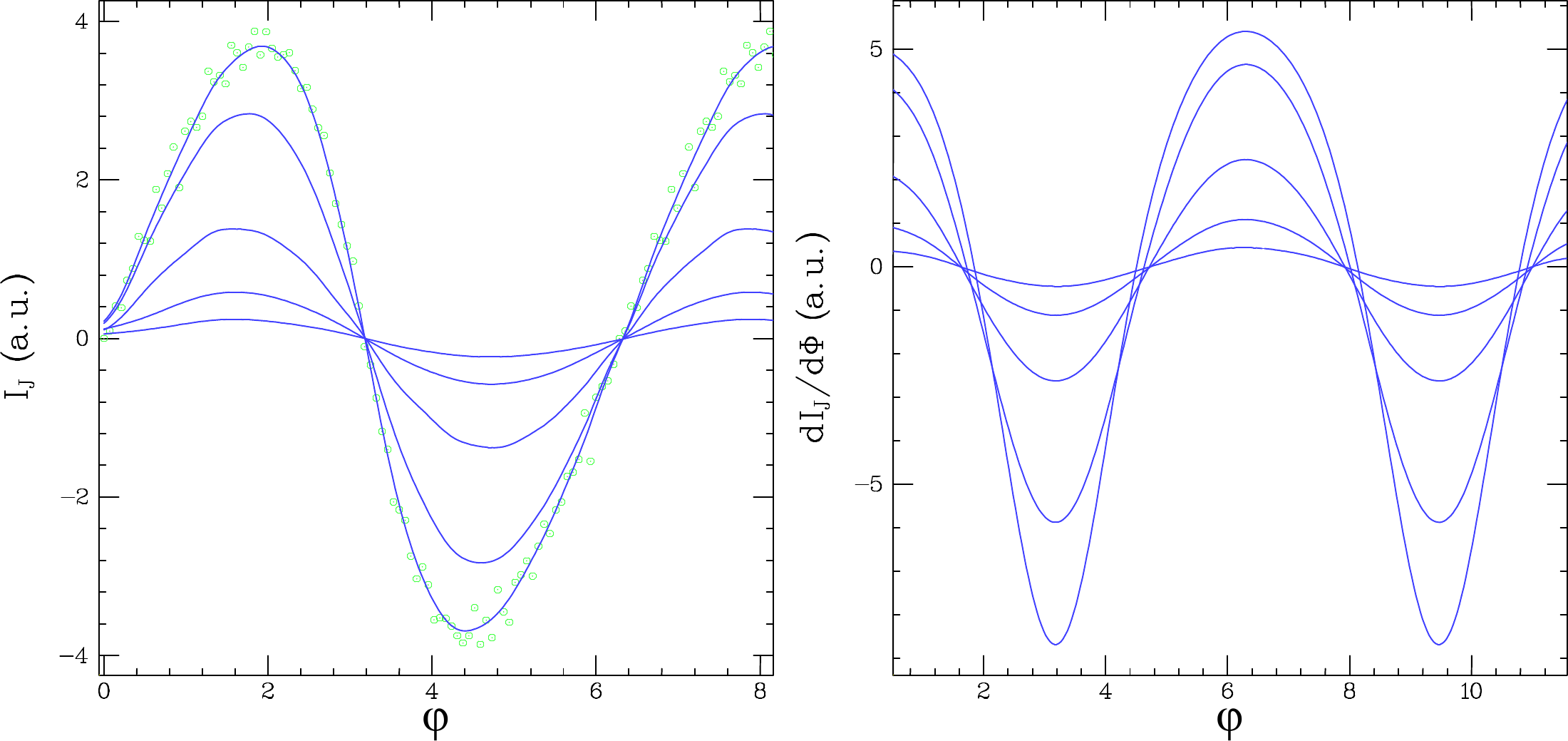}
\caption{Phase dependent Josephson current and susceptibility calculated from the Andreev spectrum shown in Fig.1 right. The temperatures correspond to 0.01,0.02,0.04,0.06 and 0.08 in the units of the superconducting gap $\Delta$. The amplitude of the minigap is estimated to be 0.04 $\Delta$. The anharmonicity is best revealed on the derivative $dI_J/d\phi$.}
    \label{IJT}
\end{figure}

Typical flux dependent spectra  obtained upon diagonalisation of the Hamiltonian   $\mathcal{H}$ (\ref{bogo}) are shown in Fig.\ref{spectre}. At energy well below the superconducting gap, energy levels   exhibit a  mean level spacing $\delta_N =E_F/N^N$  characteristic of the normal part and a $\Phi_0=h/2e$  periodicity. These constitute the Andreev spectrum.  A denser spectrum is observed above the gap with the periodicity  h/e  as expected for a normal ring, see Fig.1a and 1b. By construction, the spectrum  is perfectly symmetric  with respect to the Fermi energy.  We  observe disorder dependent fluctuations (Fig.1c)  of the  position of the energy levels  in the spectrum.   At low energy, the amplitude of these fluctuations is of the order of the mean level spacing $\delta_N $ in the N part of the ring in which Andreev levels  are confined. The flux dependent minigap  closes linearly  at $\pm \pi$ in the limit of  a very dense spectrum and can be well described as expected by $E_g(\varphi) =E_g(0) |\cos(\varphi/2)|$ \cite {spivak} (Fig.1 d).  In short junctions, this closing of the gap at $\varphi=\pi$ is directly related to the existence of conductance channels of transmission  one in a large diffusive system \cite{dorokhov} \cite{glazman}.  In long junctions the same qualitative behavior is observed even though  Andreev levels and  eigenvalues of the transmission matrix  are not simply related and that the amplitude of the minigap is much smaller than  the superconducting gap.  
 As shown in Fig.\ref {IJT}, the flux dependence of the  Josephson  current $I_J(\varphi)$ and its flux derivative %(eq.\ref{eqIJ})  
at low temperature %bares
 are sensitive to the anharmonicity of the flux dependence of  low energy levels and exhibit a slight skewness. $I_J(\varphi)$   becomes sinusoidal at temperatures larger than the Thouless energy (of the order of $0.03 \Delta$ in the simulations) according to   \cite{heikkila}. We will see in the following that the ac current response is much more sensitive than the Josephson current to the strong anharmonicity of the flux dependent minigap, and exhibits strong anomalies at $\pi$ which survive at  temperatures larger than the Thouless energy.

\section{Finite frequency linear response}

  We investigate the  linear dynamics of the  NS ring  excited by an ocillating flux $\delta\Phi(t)= \delta \Phi \exp(-i\omega t)$  leading to the time dependent  Hamiltonian $H(t)= H_0 - {\bf J}\delta\Phi(t)$ where ${\bf J}$ is the current operator. Inspired by previous work on the dynamics of persistent currents in normal mesoscopic Aharonov Bohm rings \cite{trivedi,reulet} we use as a starting point  the master equation describing the relaxation of the density matrix towards equilibrium:
\begin{equation}
\partial \rho(t)/\partial (t)=(1/i\hbar) \left[H (t),\rho \right] -\Gamma[\rho(t) -\rho_{eq}(t)]
\end{equation}
where  the equilibrium density matrix $\rho_{eq}(t)= \exp -H(t)/k_BT$  and the phenomenological relaxation tensor $\Gamma$ describes the coupling of the system to a thermal reservoir. The diagonal elements $ \gamma_{nn}=\gamma_D =\hbar/\tau_{in}$ describe the relaxation of the populations $f_n$ of the Andreev states  due to inelastic scattering such as electron-phonon or electron-electron collisions. 
Non diagonal elements $\gamma_{nm}$ describe the relaxation of the coherences $\rho_{nm}(t)$ due to interlevel transitions. We will mostly consider the limit where $ \omega$ and $ k_BT \gg \gamma_{nm} \gg \delta _N$,  for which the response function is independent of  the values of $\gamma_{nm}$. Following \cite{trivedi, reulet}, 
the  linear current response $\delta I(t) = Tr(J\delta\rho(t))+ Tr(\delta J(t) \rho_0)$  
is expressed via the complex susceptibility $\chi(\omega)= \delta I(t)/\delta\Phi(t)$,  ($\rho_0= \sum_{n} f_n (\Phi_{dc}) |n> <n| $ is the unperturbed matrix density), leading to:

\begin{equation}
\begin{array}{l}
\chi (\omega)= \displaystyle-N\frac{e^2}{2mL^2} - \sum_{n}\displaystyle\frac{\partial f_n}{\partial\epsilon_n }|J_{nn}|^2\frac{\gamma_D}{ \gamma_D-i\omega}\\
-\sum_{n,m \neq n} |J_{nm}|^2 \displaystyle\frac{f_n  -f_m }{\epsilon_n-\epsilon_m} \frac{i(\epsilon_n-\epsilon_m)  +\hbar\gamma_{nm}}{i(\epsilon_n-\epsilon_m)-i\hbar\omega  +\hbar\gamma_{nm}}
\end{array}
\end{equation} 
$J_{nm}$ is the matrix element of the current operator between the eigenstates n and m of the unperturbed Hamiltonian $H_0$ and $J_{nn}= i_n$. Using the sum rule derived from the second order perturbation of H with respect to the perturbation $J\delta\varphi$ \cite{trivedi,montambaux}:

\begin{equation}
\sum_{m \neq n}   \frac{ |J_{nm}|^2}{(\epsilon_n-\epsilon_m)} =-\frac{1}{2}\frac{\partial i_n}{ \partial \Phi} -\frac{e^2}{2mL^2} 
\end{equation}

$\chi (\omega)$ can be expressed as:

\begin{equation}
\begin{array}{l}
\chi(\omega)=\displaystyle \frac{\partial I_J}{\partial \Phi} - \sum_{n}i_n^2\frac{\partial f_n}{\partial\epsilon_n} \frac{i\omega }{\gamma_D-i\omega}-\\
\sum_{n,m \neq n} |J_{nm}|^2\displaystyle \frac{f_n  -f_m }{\epsilon_n-\epsilon_m} \frac{i\hbar\omega}{i(\epsilon_n-\epsilon_m)-i\hbar\omega  +\hbar\gamma_{nm}}
\end{array}
\label{eqchi}
\end{equation} 
This second expression clearly yields the zero frequency limit of the susceptibility 
$\chi(0)=\partial I_J/ \partial \Phi$. This expression also emphasizes the two  relaxation processes that cause frequency dependent effects as discussed in the next sections.
 
 \section{ Diagonal susceptibility and relaxation of Andreev levels  populations. }
\begin{figure}
    \includegraphics[clip=true,width=9cm]{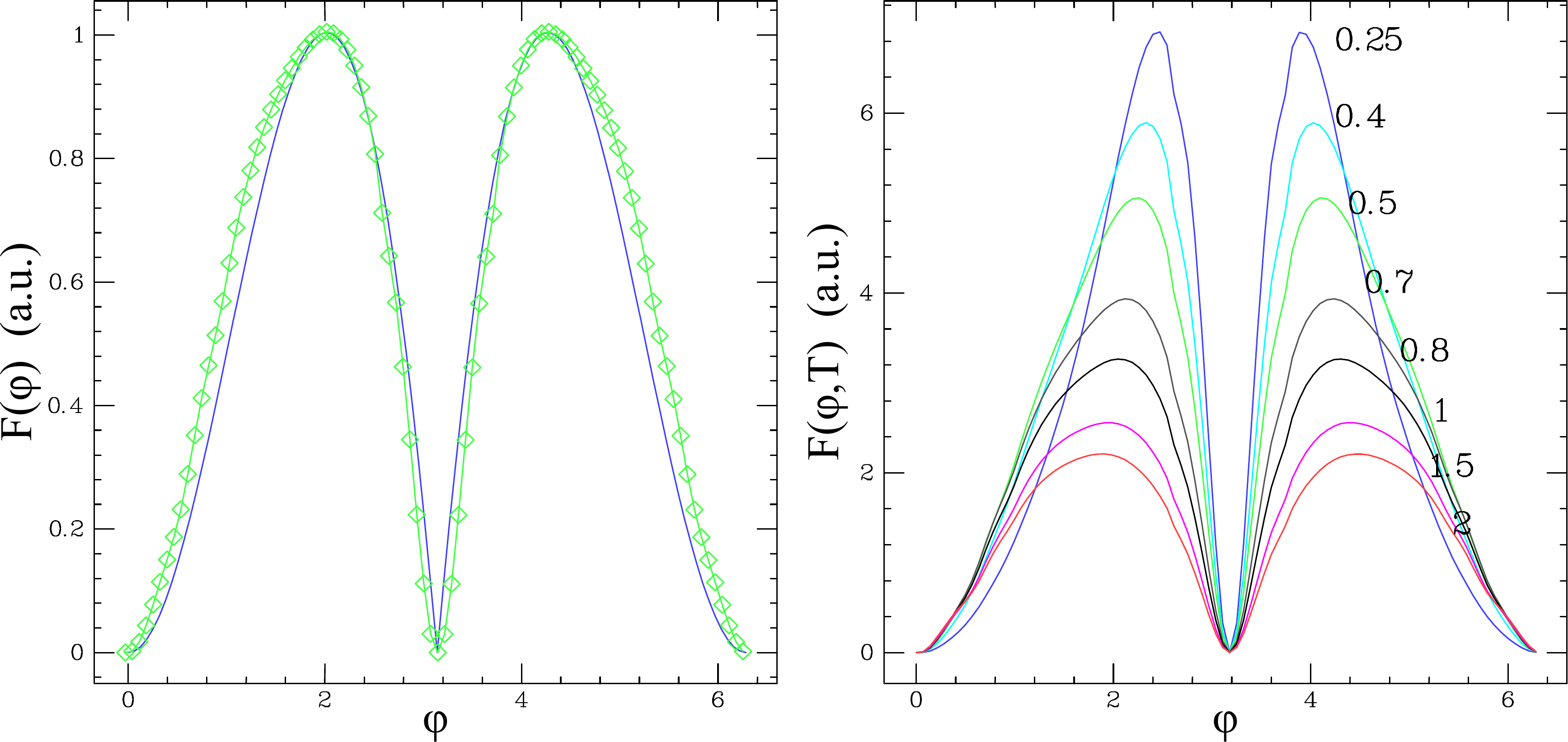}
\caption{ Right:phase dependence of the  function F computed for different temperatures in units of $\Delta$ increasing from the top to the bottom curves. Left: comparison  of the numerical results (diamonds) with  the analytical expression \ref{eqlimpitsky} (continuous line) at a temperature equal to the minigap $0.07\Delta$.}
    \label{limpitsky}
\end{figure}
                                                 
 We discuss in the following the second term of expression \ref{eqchi} that we call $\chi_D$ and is the finite frequency  non-adiabatic contribution   due to  the thermal relaxation of the populations  $f_n$ of the Andreev levels with the characteristic inelastic time $\tau_{in}$ \cite{kulikintau}.  As pointed out in \cite{trivedi}, this term  contains exclusively diagonal elements  of the current operator and is, like $\chi_{J}$, non zero only in the Aharonov Bohm ring  geometry. It is associated  to the existence of  a finite  persistent current in  a phase coherent  ring at equilibrium.  It  is proportional to the sum over an energy range $k_BT$   around the Fermi energy of the  square of the  single level current $i_n^2$.   We recast $\chi_D$ into a product of a frequency dependent term and a phase dependent one:
 \begin{equation}
\displaystyle \chi _{D}(\omega)=  \frac{i\omega\tau_{\rm in}}{1-i\omega\tau_{\rm in}} F(\varphi,T)
\label{eqchi_in}
\end{equation}
where  $F(\varphi,T)= -\Sigma_n \left[ i_n^2 \frac{\partial f_n}{\partial\epsilon_n } \right]$.
  We have  numerically evaluated  this function deriving $i_n$ from the phase   derivative of each  eigenenergy pictured in Fig.1. $F(\varphi)$  is shown   for different temperatures in Fig.\ref{limpitsky}.   As expected, $F(\varphi)$ has  a strong second harmonics component and exhibits sharp anomalies in the vicinity of  odd multiples of $\pi$ for which the minigap closes. In the continuous spectrum limit and for $k_BT  \gg E_{Th}$, $F(\varphi,T)$ can be written  in terms of the spectral current $J(\epsilon)$ and the density $n(\epsilon)$ of  Andreev levels as $ F(\varphi,T)= \int J^2(\varphi,\epsilon)/\left[k_BT n(\epsilon))\right]d\epsilon$. This function, initially introduced by Lempitsky \cite{lempitsky} to describe non equilibrium  effects in voltage biased SNS junctions, can be approximated by the analytical expression derived from Usadel equations  in the limit where $k_BT\gg E_g(0)$ \cite{ virtanen, chiodi2011}:
   \begin{equation}
   \begin{array}{l}
F_U(\varphi,T)\propto \displaystyle(\frac{1}{k_BT})\\\left[ \left[-\pi + (\pi + \varphi)[2 \pi]\right]\sin (\varphi) - \displaystyle\frac{|\sin (\varphi)|}{\pi}\sin ^2(\varphi/2)\right].
 \end{array}
 \label{eqlimpitsky}
 \end{equation}
As shown on Fig.\ref {limpitsky} this analytical form describes well the  phase dependence of the numerical results at temperatures larger than $E_{g}(0)$.  We find  however that the $1/T$  decrease at large temperature predicted in Eq.\ref{eqlimpitsky} is only   qualitatively obeyed for numerical results. %This  flux dependence reproduces also very well the experimental results  in the range of parameters $ \gamma_D\simeq \omega \ll E_g$  as shown in \cite{dassonneville} where $\chi_D$ constitutes the dominant non adiabatic response of the NS ring. 
As  pointed out in the context of atomic point contacts \cite{averin}, the dissipative component of $\chi_D$  is related via  the fluctuation dissipation theorem to the existence of a non intuitive supercurrent low frequency thermal noise \cite{averin}.  This  low frequency noise due to the closing of the minigap at $\pi$ does not exist in ordinary tunnel Josephson junctions \cite{scalapino}.
One can associate to this dissipative response an effective  phase dependent conductance $\delta G_{eff}(\varphi) = \chi''_D(\varphi) /\omega$.  The amplitude of  $\delta G_{eff}(\varphi)$ at frequencies  smaller than $\gamma_D$ and temperatures of the order or larger than $E_g$ is of the order $G_N E_g^2/( k_BT\hbar \gamma_D)$ and can be much larger than $G_N$, the normal state conductance. This component $\chi_D(\omega,\varphi)$ was recently experimentally measured on a mesoscopic NS ring \cite{dassonneville} with a very good quantitative agreement with expressions  (\ref{eqchi_in}) and (\ref{eqlimpitsky}).
% This function exhibits a sharp anomaly at $\varphi= \pi$ related to the closing of the minigap and when $\omega\tau_{\rm in} \gg 1$ this anomaly  contributes substantially   to $\chi(\varphi)$, reproducing our experimental findings  at temperatures larger than the minigap \cite{chiodi11}. 
%This expression is valid in the limit where $k_B T \gg E_{Th}$ and $\hbar \omega \ll E_{Th}$ %and $J(\varphi,\epsilon)$ and $\rho(\varphi,\epsilon)$ can be calculated using Usadel equation %[16,17]. In this limit Eqs.~(4)-(5) agree with numerical solutions of the full Usadel %equations in the presence of microwaves for the in-phase susceptibility $\chi'$ [18]. This non adiabatic contribution also gives rise to a frequency dependent absorption peaked at frequencies around $1/\tau_{in}$ which flux dependence is given by $F(\Phi,T)$.

%\begin{figure}
    %\includegraphics[clip=true,width=8cm]{chitot.pdf}
%\caption{ Flux dependence of the diagonal contribution of the susceptibility $\chi_{D}$ in the limit where $\omega \gg 1/\tau{_in}$. This quantity is the sum of the static response $\chi_J$ and the non adiabatic contribution $-F(\Phi,T)$.} 
    %\label{chitot}
%\end{figure}

\section{Nondiagonal susceptibility and microwave induced transitions in the Andreev spectrum}
\subsection{Analytical considerations}
 We  now consider the contributions of  non diagonal elements of the current operator which  describe the physics of microwave induced transitions within the Andreev spectrum. 
  \begin{equation}
 \chi_{ND}=\sum_{n,m \neq n} |J_{nm}|^2\displaystyle \frac{f_n  -f_m }{\epsilon_n-\epsilon_m} \frac{i\hbar\omega}{i(\epsilon_n-\epsilon_m)-i\hbar\omega  +\hbar\gamma_{ND}}
  \label{eqNDdisc}
 \end{equation}
 where we have assumed that all $ \gamma_{nm}$ are identical  given by a single $\gamma_{ND}$.
 
 In the continuous spectrum limit,  the average level spacing $ \delta_N$ is much smaller than the energy scales   $ \gamma_{ND}$, $k_BT$ and  $\hbar \omega$, so that  one can write:
 
 \begin{equation}
 \begin{array}{l}
 \chi_{ND} =-\int_{-E_M} ^{E_M} |J_{\epsilon,\epsilon'}|^2 \\ \displaystyle\frac{f(\epsilon) - f(\epsilon' )}{\epsilon-\epsilon'} \frac{i\hbar\omega}{i(\epsilon-\epsilon')-i\hbar\omega  +\gamma_{ND}}n(\epsilon)n(\epsilon')d\epsilon d\epsilon'
  \end{array}{}                                 
  \label{eqND}
 \end{equation}
where $E_M$ is a high energy cutoff of the order of the bandwidth, from now on arbitrarily taken as unity, and $n(\epsilon)$ is the density of states at energy
$\epsilon$. In the limit where the induced minigap is very small compared to the superconducting gap $\Delta$  (long junction)   one can approximate the density of states  as a step function at $E_g(\varphi)$:  $n(\epsilon,\varphi)= n_0\left[ \theta(\epsilon-E_g(\varphi)) +\theta(-\epsilon-E_g(\varphi))\right]$,  (with $\theta (x)$,  the Heaviside function ). In the next paragraph we also assume that $|J_{\epsilon,\epsilon'}|^2$ %depends only on the energy difference $\epsilon -\epsilon'$ on an energy scale which is large compared to all other energy scales of the problem and will %
can be approximated by a constant $J^2$. We will see that this approximation is  valid when  %both the energy sum and difference $\epsilon +\epsilon'$, $\epsilon +\epsilon'$ are of the order or larger than the minigap.  This corresponds to the regime where
 $k_BT \ll E_g  < \hbar\omega $ where the dominant contribution comes from matrix elements nearly independent of $\varphi$.
This leads to:  
  \begin{equation}
 \begin{array}{l}
 \chi_{ND} =-n_0^2 \displaystyle\int \int_{|\epsilon|,|\epsilon'|\geq E_g (\varphi)}d\epsilon d\epsilon'  \\ \left[|J|^2 \displaystyle\frac{f(\epsilon) - f(\epsilon' )}{\epsilon-\epsilon'} \frac{i\hbar\omega}{i(\epsilon-\epsilon')-i\hbar\omega  +\gamma_{ND}}\right]
  \end{array}{}
  \label{eqNDap}
 \end{equation}

\begin{figure}    
    \includegraphics[clip=true,width=9cm]{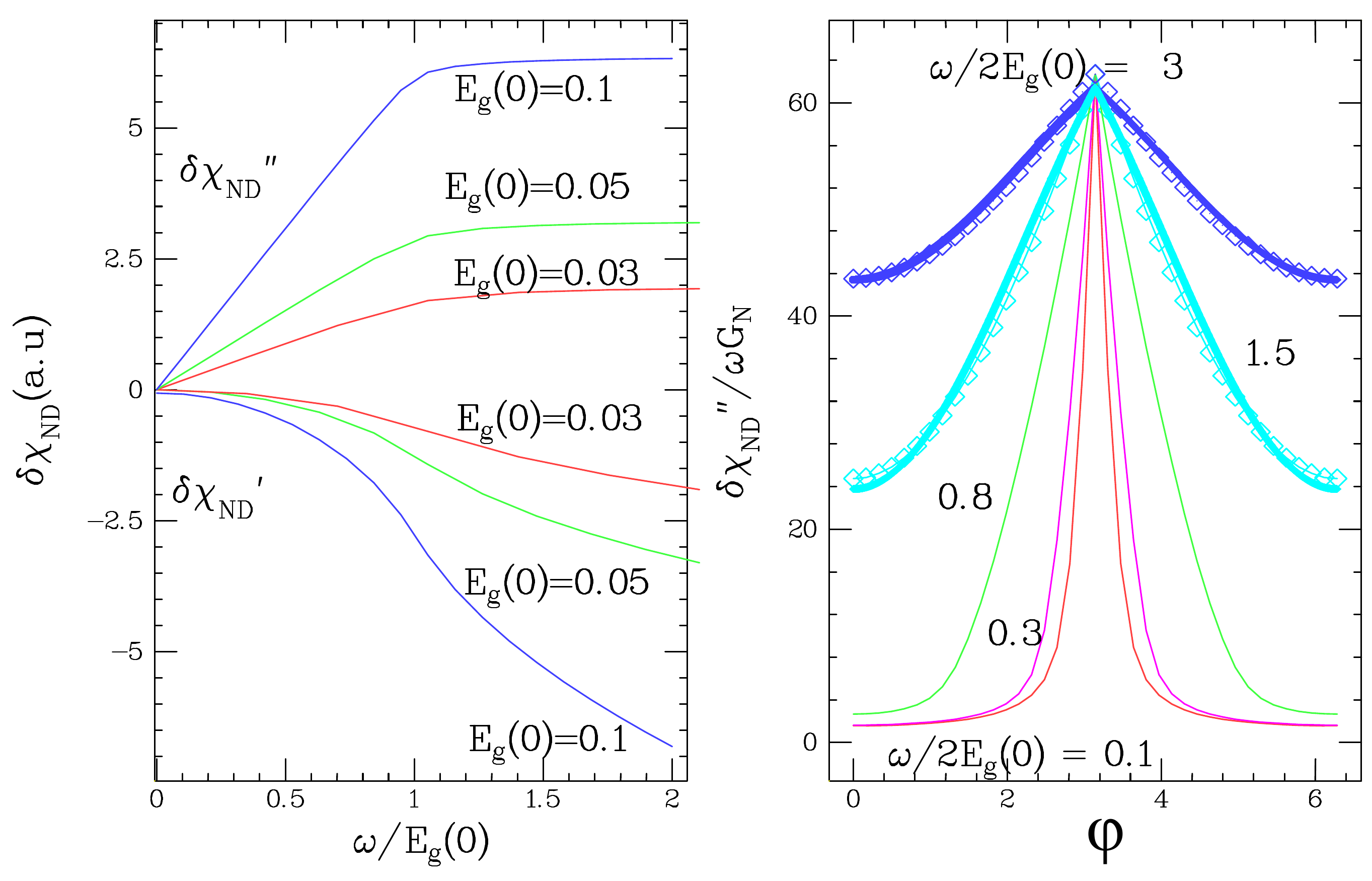}
\caption{ Non diagonal susceptibility calculated assuming no phase dependence for the  non diagonal matrix elements of the current operator. The  temperatures and frequencies investigated correspond to $T \ll \hbar\omega$.  The values of $\gamma_{ND}$ and $ k_BT$ were both taken  equal to $0.01$ i.e. much smaller than the minigap $2E_g(0)$. Left: frequency dependence of  $\delta \chi''_{ND}$ and $\delta \chi'_{ND}$ dissipative and non dissipative responses for different values of the minigap. Right: phase dependence of $\chi''_{ND}$ for different frequencies. The  thick continuous lines corresponds to a fit with a $-|\cos(\varphi/2)| \propto- E_g(\varphi) $ dependence.}
\label{chindsansj2}
\end{figure}

We define $\delta \chi'_{ND}= \chi'_{ND}(\pi) -\chi'_{ND}(0)$ and $\delta \chi''_{ND}= \chi''_{ND}(\pi) -\chi''_{ND}(0)$ as the  amplitudes  of the flux dependent components of the real and imaginary parts of $\chi_{ND}(\Phi,\omega)$. The frequency dependence of these quantities are depicted  in Fig.\ref{chindsansj2} for several values of the minigap larger than the temperature. We find that  $\delta \chi'_{ND}$  is negative and  decreases slowly at low frequency with an inflexion point  at $\omega= E_g(0)/\hbar$; $\delta \chi''_{ND}$  is positive and increases linearly  with frequency up to $\omega= E_g(0)/\hbar$  and is independent of frequency at larger values. These results, in agreement with Kramers Kronig relations, 
show that the minigap is  the fundamental frequency scale for $\chi_{ND}(\varphi)$.  In the limit where $\gamma_{ND} \ll \omega$ and  $\gamma_{ND} \ll k_BT $ ,   $ \gamma_{ND}/\left[(\epsilon-\epsilon' -\hbar\omega)^2+\gamma_{ND}^2)\right]$ entering in $\chi''$ deduced from  Eq.\ref{eqNDap} can be approximated by the delta function: $\delta(\epsilon-\epsilon' -\hbar\omega)$. It is  then possible to express simply   $\chi''_{ND}(\omega,\varphi)$ analytically as:
 \begin{equation}
 \chi''_{ND} = n_0^2 |J|^2 \int_{|\epsilon|\geq E_g (\varphi)} \left[f(\epsilon) - f(\epsilon +\hbar \omega)\right]d\epsilon
 \label{eqND2}
 \end{equation}       
 Because the variation in $\varphi $ is only contained in the integration limits, we find that in the frequency range where  $\omega\gg k_BT$   $\chi''(\varphi)$  mimics the minigap (with a minus sign)  in the  flux domain  where $ \hbar\omega\geq 2E_g(\varphi)$ and reads $ \chi ''_{ND}(\varphi,\omega) = G_N( \omega -2 E_g(\Phi)/ \hbar)$. The normal state conductance $G_N = \chi ''_{ND}(\pi)/\omega$ (where the minigap closes) can be expressed as $G_N = |J|^2n_0^2$. On the other hand  at low frequencies below  $E_g(\varphi)$, $\chi ''_{ND}(\varphi)$ is  equal to zero. As a result when $\omega \ll E_g(0)$  the flux dependent absorption exhibit sharp peaks at odd multiples of $\pi$ which amplitude scales linearly with $\omega$ as shown on Fig.\ref{chindsansj2}. One finds that at low frequency the ratio $\delta \chi''_{ND}/ \delta \chi_J = \delta \chi''/\delta \chi_J $ varies like $\hbar\omega /E_g$. There is however no  simple  analytical expression for the  complete phase and frequency dependences of $\chi'(\omega,\varphi)$  owing to the fact that according to  Eq.\ref{eqNDap},  it explicitly depends logarithmically   on the energy cutoff $E_M$.

%At this stage it would be  tempting to  extend the analysis  of Eq.\ref{eqND2} 
 In the opposite limit  of high temperature $T \gg  E_g \gg \hbar\omega$  we  can easily find  from Eq.\ref{eqNDap} that the ratio $\delta \chi''_{ND}/  \delta \chi'_J(T) $ varies like $\hbar\omega/k_BT$.  It is however not possible to use Eq.\ref{eqND2} to deduce the phase dependence of $ \chi'' _{ND} $.  This equation   relies on a crude approximation neglecting the phase dependence of the non diagonal matrix elements of the current operator.  We will show in the next paragraph devoted to numerical calculations, that this approximation  is  only reasonable at low temperature and large frequency where, in the expression of $\chi''_{ND}$,  only a small number of matrix elements contribute.  These are matrix elements $|J_{\epsilon,\epsilon'}|^2$  coupling  negative energy levels close to the minigap   to positive energy levels much larger than $E_g$. These matrix elements have indeed only a very small phase dependence. On the other hand, at high temperature,   $k_B T \gg E_g$,   a large number of matrix elements $|J_{\epsilon,\epsilon'}|^2$ contribute to the integral   in $\epsilon'$ in  Eq. \ref{eqND}. 
	%Therefore, in this limit  $T \gg \hbar\omega \sim eq E_g(0)$, the phase dependence of 	$\chi''_{ND}$ is then determined by the sum of a large number of non diagonal  matrix elements $|J_{nm}(\varphi)|^2$ with $m\neq n$. 
		%whose phase  is therefore  opposite in sign to the variation of $F(\varphi) = \sum_n |J_{nn}|^2$.% The phase dependence of $\chi_{ND}$%
 We can  then estimate their contribution to the phase dependence  of  $\chi''_{ND}$ using  the fact that $Tr(J^2)=  \Sigma_n |J_{nn}|^2+\Sigma_{n,m \neq n} |J_{nm}|^2$  does not depend  on the Aharonov-Bohm phase  just like  $ Tr(\mathcal{H})$, (since the Aharonov Bohm phase only affects non diagonal matrix elements of $\mathcal{H}$). The sum  of all non diagonal matrix elements $|J_{nm}(\varphi)|^2$ with $m\neq n$ is thus   opposite in sign to the variation of $F(\varphi) \propto \sum_n |J_{nn}|^2$ at large $T$.	
 Therefore,  in the limit  $T \gg \hbar\omega \simeq E_g(0)$, where the sum of a large number of non diagonal  matrix elements $|J_{nm}(\varphi)|^2$ with $m\neq n$  contribute to the phase dependence of $\chi''_{ND}$,  {\bf the  phase dependence of $\chi''_D$ and $\chi''_{ND}$ are thus expected to be reversed from one another}.   Results of numerical simulations presented in the next paragraph  agree with this simple  qualitative prediction. 

\subsection{Numerical results for the non diagonal susceptibility.}

The  non-diagonal matrix  elements of the current operator $\vec J = (\hbar/i) \vec\nabla -q \vec A$  along the ring are calculated from the eigen wavefunctions according to: 
\begin{equation}
\begin{array}{l}
J_{nm} = \frac{\hbar}{i} \sum_j \Psi_n^{e*}(x_j, y_j) (\Psi_m^e (x_j+1, y_j)- \Psi_m^e(x_j, y_j) +eA(x_j)) \\+ \Psi_n^{h*}(x_j, y_j) (\Psi_m^h (x_j+1, y_j)- \Psi_m^h(x_j, y_j) -eA(x_j)).
\end{array}
\end{equation}

where $\Psi_m^e(x_j, y_j)$ and $\Psi_m^h(x_j, y_j)$ correspond respectively to the electron and hole  components of the wave function at  point $j$ of  coordinates ($x_j$, $y_j$) in units of $a$.  
 The phase dependence of the square modulus of these matrix elements is shown in Fig.5 for various indexes n and m  on the same side (a)  or  on either sides (b)  of the minigap.  The index n and m  are  taken respectively  positive above and negative below the minigap. Whereas $|J_{-11}(\varphi)|^2= i_1^2(\varphi)$ exhibits a strong peak at $\varphi=\pi$ %which amplitude is twice the dip in the diagonal matrix element $|J_{11}(\varphi)|^2$),
 the amplitude of $|J_{-1n}(\varphi)|^2$ is much smaller at large n with a phase dependence  that is smooth around $\pi$ and  a maximum around zero phase. On the other hand matrix elements  $|J_{-nn}(\varphi)|^2$ corresponding to states symmetric with respect to the minigap, i.e. electron hole symmetric states, keep a phase dependence  peaked at $\pi$ similar  but reversed in sign compared to $|J_{1,1}|^2(\varphi)$.  Their amplitude decreases only slowly with n in contrast to the fast   amplitude decrease of the diagonal matrix elements $J_{nn} =i_n$. This difference between the phase dependence of $|J_{-nn}(\varphi)|^2$ compared to $|J_{-1n}(\varphi)|^2$  can qualitatively explain the evolution of the shape of $\chi_{ND}(\varphi)$ in the limit $\omega\gg k_BT$ compared to  $\omega\ll k_BT$. In the first case, the main contribution stems from matrix elements $|J_{-1n}(\varphi)|^2$ where $n\gg 1$  with a very small phase dependence, whereas  in the second case, a much larger number of matrix elements contribute to $\chi''_{ND}$, including the electron hole  symmetrical ones $|J_{-nn}(\varphi)|^2$.

  $\chi''_{ND}(\varphi)$ is computed  from these matrix elements and the related energy spectrum following Eq.\ref{eqNDdisc}. We took $\gamma_{ND} =3\delta_N $ in order to reproduce the continuous spectrum limit.  The results concerning the imaginary component $\chi''_{ND}(\varphi)$ are shown in  Fig.7 a and b for $\hbar\omega > k_BT$ and $\hbar\omega < k_BT$ respectively.  In the first case  $\hbar\omega > k_BT$, we find  good qualitative agreement with our analytical findings  neglecting the flux  dependence of the $|J_{nm}|^2$ in particular  $\delta\chi''_{ND}(\varphi)$  is peaked at $\pi$ and its amplitude increases linearly with frequency up to $\hbar\omega= 2E_g$, whereas in the second case  $\hbar\omega < k_BT$, we find that the  shape of $\delta\chi''_{ND}(\varphi)$ is  very similar  to the opposite of the function $F(\varphi)$ , (giving the phase dependence of the average square of the single level current), with a characteristic bump at $\varphi=0$  ( Eq.\ref{eqchi_in}). A similar behavior is  found for $\delta\chi'_{ND}(\varphi)$.
	
	\begin{figure}
   \includegraphics[clip=true,width=9cm]{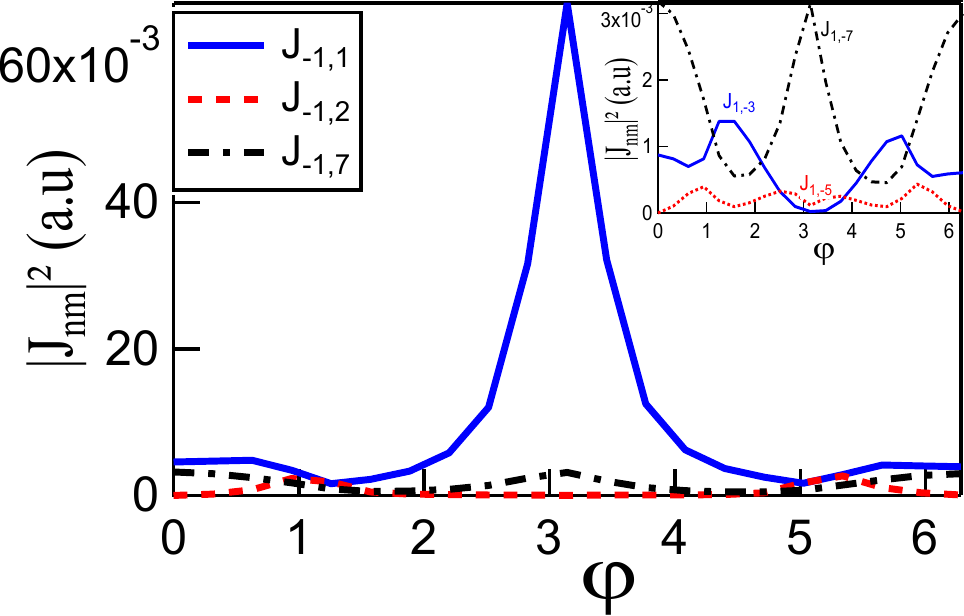}
\caption{Phase dependence of the  non diagonal current matrix elements   $|J_{-1,n}|^2$ coupling the highest level below the minigap to levels above the minigap. Inset: zoom on  $|J_{-1,n}|^2$  with $n > 1$ which have a very small phase dependence compared to $|J_{-1,1}|^2$ . They are  obtained from the exact diagonalisation of the spectrum of an NS ring whose  normal region size is $90\times 30 $ and $W/t=2$.
The minigap amplitude is $2E_g(0) =8 \delta_N$. }
    \label{J2}
\end{figure}
\begin{figure}
   \includegraphics[clip=true,width=9cm]{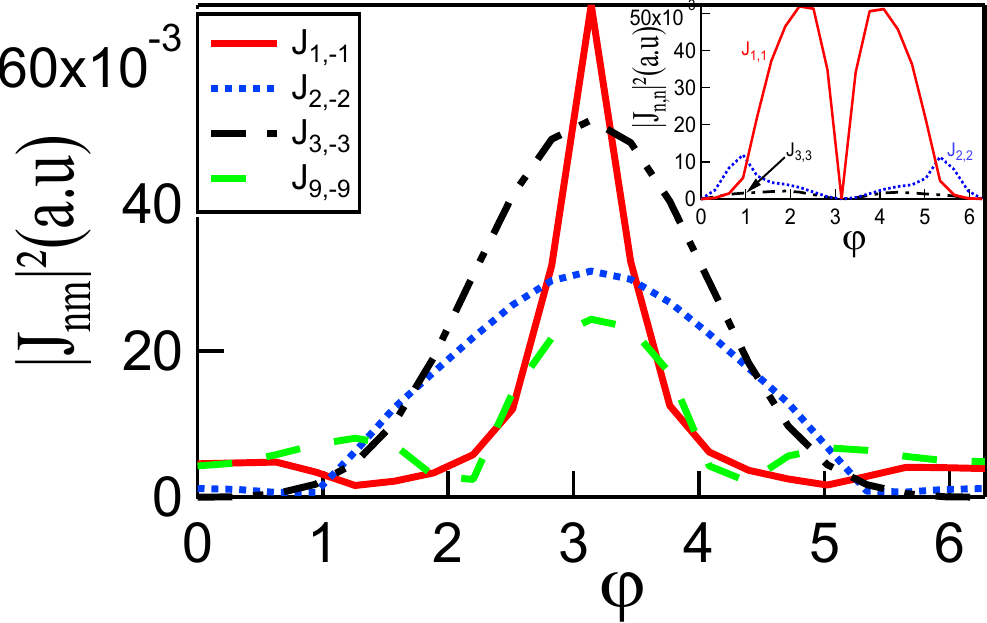}
\caption{Phase dependences of the  electron hole symmetrical non diagonal matrix elements  compared to the diagonal ones obtained from the exact diagonalisation of the spectrum of an NS ring whose  normal region size is $90\times 30 $ and $W/t=2$.
The minigap amplitude is $2E_g(0) =8 \delta_N$. }
    \label{J2}
\end{figure}
\begin{figure}
 \includegraphics[clip=true,width=8cm]{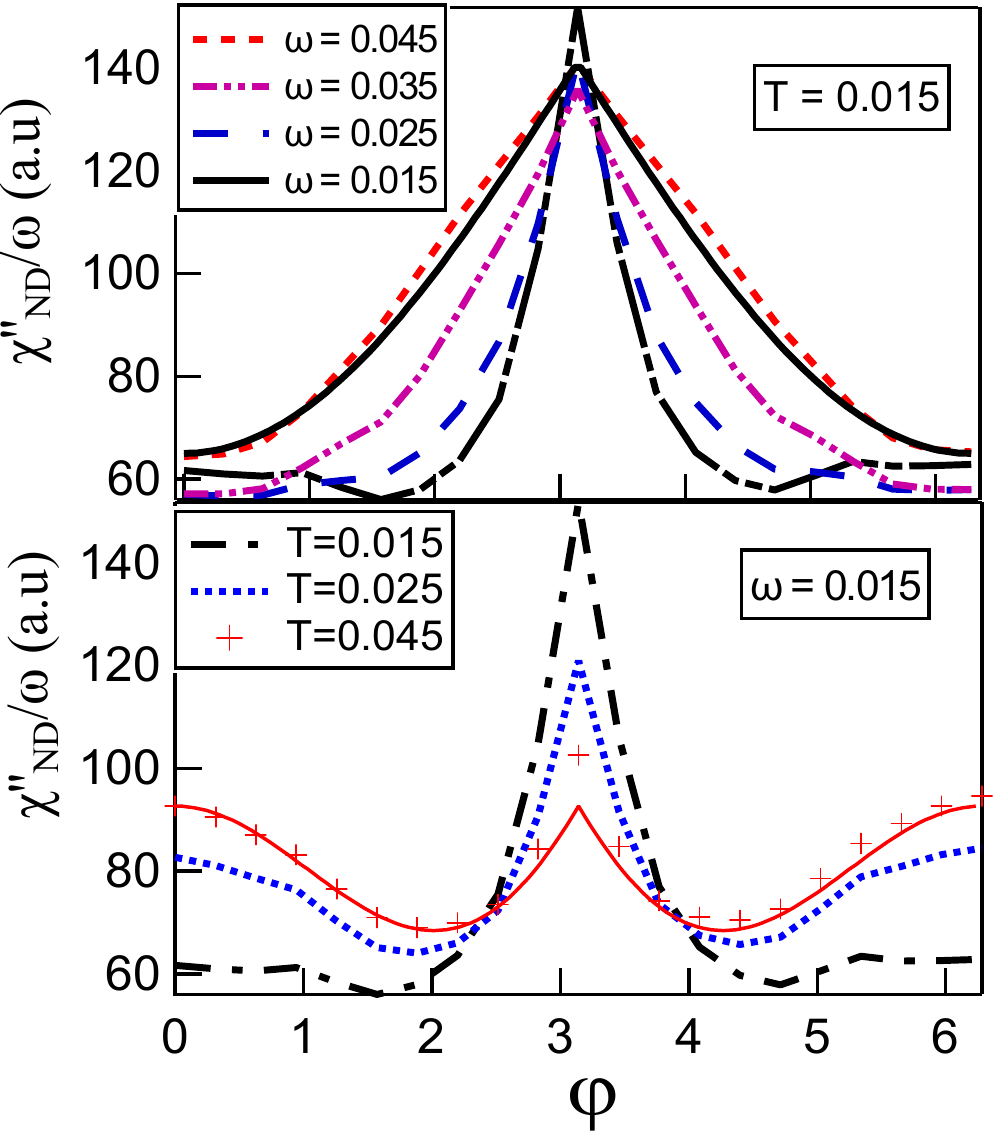}

\caption{Evolution of  the phase dependence of $G_{ND}=\chi''_{ND}/\omega$ obtained from the exact diagonalisation of the spectrum of an NS
 ring (size $90\times 30 $) $W/t=2$ using Eq.\ref{eqNDdisc} 
$\gamma_{ND}=3\delta_N$: Top panel:$k_BT=2\delta _N $ and  different frequencies $ \hbar\omega > k_BT$ below and of the order of the minigap, note the good agreement with the data obtained Fig.4, neglecting the phase dependence of  the current matrix elements. Bottom panel: $\hbar\omega =2\delta_N $ and  different temperatures  $ k_BT > \hbar\omega $ below and of the order of the minigap. For the largest temperature the phase dependence observed is close to the opposite of the function $F_U(\varphi)$, continuous line.}
    \label{gndnum}
\end{figure}

\begin{figure}
   \includegraphics[clip=true,width=8cm]{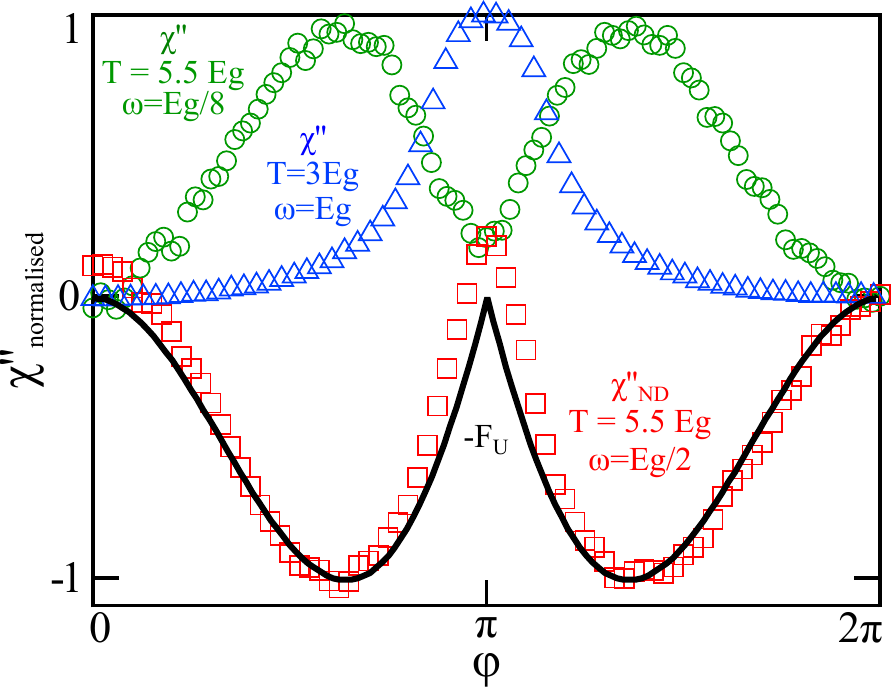}
\caption{The phase dependent dissipative response measured experimentally \cite{dassonneville, dassonnevillelong} is shown for  3 different regimes: circles: $\omega\tau_{in} \simeq 1$ where $\chi''_D$ is the dominant contribution with a phase dependence  following  $F(\varphi)$. Triangles: at frequencies and $k_BT$ of the order of $E_g(0)$ the dissipative response is dominated by $\chi''_{ND}$ and  we observe a phase dependence  peaked at $\pi$ which resembles the minigap. Squares: in the limit where  $k_BT > \omega \geq E_g$ a phase dependence opposite to $F(\varphi)$ is found  as expected for the contribution of the non diagonal matrix elements of the current operator, in agreement with the results in Fig.7.  The continuous line is $-F_U(\varphi)$ calculated from expression \ref{eqlimpitsky}. }
    \label{figexp}
\end{figure}

 \section{Conclusion}
 
 We have developed a simple model for the computation of the ac linear response of an NS  diffusive ring to a high frequency flux in the long junction limit. Starting from the dc phase dependent Andreev spectrum  and wave functions of the ring, we use a Kubo formula adapted for the Aharonov Bohm geometry which yields the complex susceptibility of the ring as a function of the energy levels and matrix elements of the current operator.
 We clearly identify 2 different finite frequency contributions superimposed to the dc response which is the  flux derivative of the Josephson current. The first one, expressed in terms of the diagonal element of the current operator, can be understood as the Debye relaxation of the populations of the Andreev states.  The second one, expressed in terms of the non diagonal matrix elements of the current operator, describes inter level transitions within the Andreev spectrum.  It is striking that numerical simulations  on small systems with less than 10 levels in the energy scale corresponding to the minigap can  reproduce the experiments \cite{dassonneville, dassonnevillelong} investigating  the ac susceptibility of an NS ring  where $E_g/\delta_N$ is of the  order of 1000, as illustrated in Fig.8. The phase dependent dissipative response is shown for  3 different regimes:(i) $\omega\tau_{in} \simeq 1$ where $\chi''_D$ is the dominant contribution with a phase dependence   well described  by $F(\varphi)$, (ii) at frequencies and $k_BT$ of the order of $E_g(0)$ we observe a phase dependence  peaked at $\pi$ which resembles the minigap, (iii) finally in the limit where  $k_BT > \omega \geq E_g$ a phase dependence opposite to $F_U(\varphi)$ is found,  as expected for the contribution of the non diagonal matrix elements of the current operator, in agreement with the numerical results in Fig.7.   On the other hand, whereas Usadel equations \cite{virtanen} provide an excellent agreement between the numerical and experimental findings for the diagonal contribution $\chi_D$, the high frequency regime yields different results. In particular the predicted  phase oscillations  of the susceptibility  do not reproduce our findings.
 
 We acknowledge   M. Aprili, F. Chiodi, R. Deblock, M. Feigelman, T.T. Heikkil$ \ddot{\rm a} $, A. Levy-Yeyati, G. Montambaux, C.Texier,  K. Tikhonov  and  P. Virtanen for fruitful discussions. We have benefited from financial support from the grant MASH of the French agency of research ANR.

\end{document}